# Rationalization of the π-σ (anti)aromaticity in all metal molecular clusters

Ayan Datta and Swapan K. Pati*

*Theoretical Sciences Unit and Chemistry and Physics of Materials Unit, Jawaharlal Nehru Center for Advanced Scientific Research, Jakkur P. O, Bangalore-560064, India.*

**RECEIVED DATE (automatically inserted by publisher)**; pati@jncasr.ac.in

The last decade has witnessed a rapid progress in the new chemistry of small metal clusters of Al, Si and Ga facilitated by computational strategy, synthesis and characterization.[1-3] These metal clusters have a close resemblance with the cyclic organic π-conjugated molecules in structure as well as properties. $Al_4Li_4$ and their anions, $Al_4Li_3^-$ or $Al_4^{4-}$ follow the properties of cyclobutadiene ($C_4H_4$).[4] They also undergo interaction with transition metals like Fe and Ni to form complexes of the type $(Al_4Li_4)$-$Fe(CO)_3$ and sandwich complexes of the type: $(Al_4Li_4)_2Ni$ thereby resembling $C_4H_4$ in its role as a ligand.[5] $Al_4^{4-}$ also satisfies the simple Huckel criteria for antiaromaticity as it possess 4π electrons in its frontier π-orbitals. Despite these similarities between these all-metal systems and their organic counterpart ($C_4H_4$), serious doubts have recently been raised whether these systems are aromatic or antiaromatic, based on π-electron scheme alone.[6-7] The σ-backbone appears to be quite important and based on nucleus independent chemical shift (NICS), Schleyer and co-workers suggested that these clusters are net aromatic. But Boldyrev, Wang and co-workers concluded that $Li_3Al_4^-$ and $Li_4Al_4$ clusters are net antiaromatic. Santos and co-workers agreed with the net antiaromaticity of these clusters on the basis of electron localization function (ELF) analysis.[8]

In this letter we show that the confusion associated with these clusters can be settled through a simple σ-π separation analysis which provides an unambiguous answer to all such questions. The different results obtained by the previous workers arise primarily because of the indirect methods used to characterize aromaticity/antiaromaticity. We stress that the σ-backbone is quite an important component in the structural features of almost all molecular systems. Even in $C_6H_6$, it has been found that the σ-backbone is responsible for the symmetric $D_{6h}$ structure and the π-electrons actually tend to distort the symmetric structure.[9-10]

We have considered a variety of molecular systems: $Al_4Li_4$, $Al_4Li_4^{2-}$, $Ga_4Li_4$, $Al_4^{2-}$ and compared with $C_4H_4$ and similar organic analogues at each step of our σ-π analysis. These systems have either 4π, 6π or 2π electrons in their frontier orbitals and provide a diverse set for studying aromaticity or antiaromaticity. All the geometries were optimized at the B3LYP/6-311G++ (d, p) level[11-12] (see supporting information file for structures and energies). The ground state geometry for both $Al_4Li_4$ and $Ga_4Li_4$ have a planar rectangular structure for the the ring with the Li ions occupying positions so as to maintain a $C_{2h}$ architecture. The bond-length alteration (BLA) for $Al_4Li_4$ and $Ga_4Li_4$ are 0.12 and 0.16 Å respectively. Note that the same for $C_4H_4$ is 0.2 Å . The fact that the BLA for $Ga_4Li_4$ is more than that in $Al_4Li_4$ suggests that $Ga_4Li_4$ is more antiaromatic than $Al_4Li_4$, in analogy with $C_4H_4$ and a σ-π separation should be ideal to quantify such a statement.

We distort the geometry optimized structures by ΔR (where ΔR is the difference between the long M-M and short M-M bond in the $M_4$ ring) so that that the distortion keeps the sum of two adjacent M-M bonds constant [Scheme 1 (a)]. The energy associated with the distortion is partitioned into σ and π components as $\Delta E_\pi = \Delta E_{GS} - \Delta E_\sigma$. One of the simplest methods to get the contribution associated with a distortion only along the σ-backbone is to freeze the π-electrons with all-parallel spins. The σ-backbone for a $M_4$ ring with 4π electrons can be modeled as $M_4^{4-}$ with a H.S configuration (S=2) with all the 4π electrons being parallel [Scheme 1 (b)]. Similarly, for the 2π electron sytems like $C_4H_4^{2+}$ and $Al_4^{2-}$, S=1 corresponds to the H.S state. For the 6πe $Al_4Li_4^{2-}$ however, there are only four π-orbitals and thus a H.S configuration with S=3 is not feasible, rather two parallel spins with S=1 state in $Al_4Li_4^{2-}$ corresponds to the H.S state. We thus define, $\Delta E_\sigma = \Delta E_{HS}$. Such an analysis gives a very clear picture of the nature of interactions in the system and has been extensively used in the literature for various organic molecules.[13-14] For the HS systems, we perform UB3LYP calculations at the same basis set level with annihilation of the first spin-contaminant.

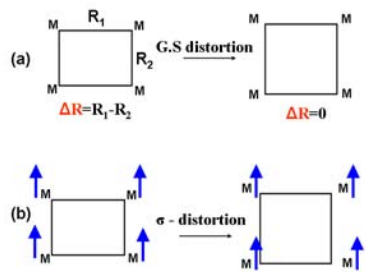

**Scheme 1.** (a) The distortion mode for the $M_4$ rings (M=C, Al, Ga) in the ground state. Li atoms not shown for the sake of clarity. (b) The distortion in the σ – electrons involving the distortion in a high spin configuration.

In Figure 1, we plot the σ-energy and the π-energy as a function of the distortion parameter, ΔR. In the inset, the core energy, $V_{core}$ (sum of kinetic energy and nuclear-electron (ne) interactions), electron-electron interactions ($V_{ee}$) and the nuclear-nuclear interactions ($V_{nn}$) are plotted. For all the systems, we find that the π-electrons have a general tendency of forming distorted structure (π-energy is most stable at large ΔR) while the σ-framework oppose the distortion and tends to equalize the bonds. The final structure and thus the aromatic/antiaromatic features will crucially depend on the predomination of either of the forces. In Figure 1(a), the result for the well-known $C_4H_4$ system is shown. The instability associated with the σ-backbone distortion is very little (4 kcal/mol for ΔR=0.1) while the stability for π-distortion is quite substantial (22 kcal/mol for ΔR=0.1), clearly overwhelming the tendency for σ-backbone equalization. Thus the $C_4H_4$ has a rectangular structure and is overall π-antiaromatic with a minor σ-aromatic component. Both $V_{ee}$ and $V_{nn}$ are destabilized with distortion while the $V_{core}$ component is stabilized. We have further analyzed that it is the $V_{ne}$ term in the $V_{core}$ that favors the distorted structure. This is easy to understand as the $V_{ne}$ component is associated with the electron-lattice interactions and lead to Jahn-Teller stabilization in the distorted structure. However components like $V_{ee}$ and $V_{nn}$ stabilizes the ΔR=0 structure associated with the delocalized π-electrons (for nonzero ΔR, the electron density is localized in shorter bonds).

For the all-metal system however, the σ-π separation energy play a crucial role. For example, in $Al_4Li_4$, the distortion in the σ-framework leads to a destabilization of 2.5 kcal/mol while the π-framework gains energy of 3.5 kcal/mol (Figure 1(b)). The ground state energy is thus stabilized by the distortion along the ring. Accordingly thus the $Al_4Li_4$ is π-antiaromatic though the σ-aromatic component is also substantial. However overall $Al_4Li_4$ is antiaromatic as the π-antiaromaticity exceeds the σ-aromaticity by 1 kcal/mol. The energy components also follow very similar trends like that for $C_4H_4$ (Fig. 1(b), inset). We derive similar conclusion for the $Ga_4Li_4$ also and the π-stabilization associated with the distortion is 4 kcal/mol while σ-destabilization is 2.5 kcal/mol (seen in Fig. 1(c)). The distorted π-antiaromatic structure is thus stabilized by an amount of 1.5 kcal/mol, 0.5 kcal/mol more than that for $Al_4Li_4$. Thus the π-antiaromaticity follows the order: $C_4H_4 > Ga_4Li_4 > Al_4Li_4$.

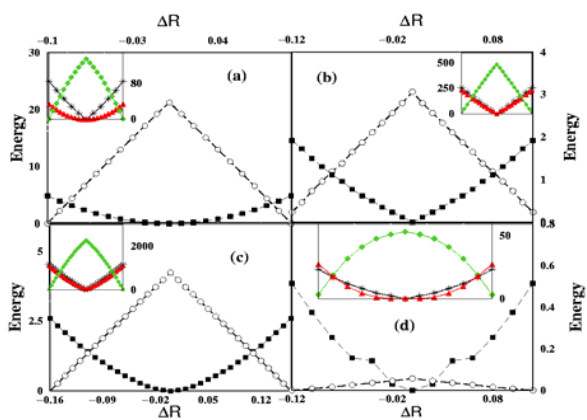

**Figure 1.** Variation of the σ-energy (square) and the π-energy (circles), both in kcal/mol as a function of the distortion axis, ΔR for (a) $C_4H_4$ (b) $Al_4Li_4$ (c) $Ga_4Li_4$ and (d) $Al_4Li_4^{2-}$ derived from $(Al_4Li_4)Fe(CO)_3$. The insets show $V_{core}$ (green), $V_{ee}$ (black) and $V_{nn}$ (red) components in the ground state structures. All the energies are scaled to make the most stable geometry zero in energy and positive values in Energy-axis correspond to destabilization.

The fact that this simple σ-π separation gives a very clear picture for the nature of aromaticity/antiaromaticity is evident from Fig. 1(d). We retrieve the structure of the $Al_4Li_4$ unit from the organometallic complex $(Al_4Li_4)$ Fe $(CO)_3$ and consider the $Al_4Li_4$ dianion system. The interaction of the $Fe(CO)_3$ unit with the $Al_4Li_4$ converts it into a 6π aromatic system with small BLA. In Fig. 1(d) we distort this dianion of $Al_4Li_4$ and perform similar analysis. Contrary to the previous cases, in $Al_4Li_4^{2-}$, the stabilization associated with the equalization of the σ-backbone overwhelms the instability due to π-electron localization by 0.5 kcal/mol and forces the system to be aromatic. This is of course true for $C_6H_6$ where σ-delocalization exceeds the π-localization by 6 kcal/mol.[15-16] In Fig. 2 (a), the energy profile is plotted for $C_4H_4^{2+}$ which shows an overwhelming π-delocalization compared to the smaller σ-localization. Similarly, for the all-metal system, $Al_4^{2-}$, the ground state corresponds to a square geometry with Al-Al bond length=2.54 Å. This is readily understood from the plot as the π-destabilization associated with the distortion exceeds the stability in the σ-backbone due to distortion (Fig. 2 (b)) though again the energy scales for the σ and π distortion are comparable. Also, as a general rule, we find that the $V_{ne}$ term favors distortion.

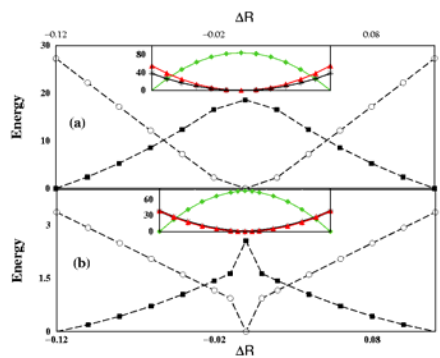

**Figure 2.** Same variation (including the inset) as in Fig. 1 for (a) $C_4H_4^{2+}$ and (b) $Al_4^{2-}$.

In summary, we have shown that all-metal molecular clusters like $Al_4Li_4$ and $Ga_4Li_4$ are predominantly π-antiaromatic although there is a significant contribution from the σ-aromaticity as well due to close proximity in σ/π energy scales compared to the equivalent organic systems. We believe that our analysis provides a tool for assignment of aromaticity/antiaromaticity in all-metal clusters.

**Acknowledgement.** SKP thanks CSIR and DST, Govt. of India for research grant.

**Supporting Information Available:** Structures, cartesian coordinates, ground state energies and Complete Ref. 11. This material (PDF) is available free of charge via internet at http://pubs.acs.org.

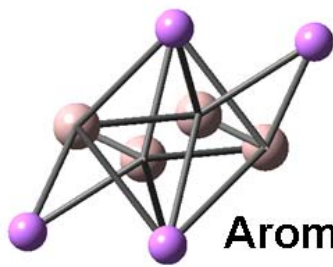
Aromatic or antiaromatic?

A σ-π separation analysis of the energies in $Al_4Li_4$ reveals that the system is more π-antiaromatic than the σ-aromaticity in it. This is true also for $C_4H_4$ and $Ga_4Li_4$. Unlike $C_4H_4$ that has a very large component of π-antiaromaticity, for these all-metal clusters, these energy scales are comparable though π-antiaromaticity is the major driving force for the distortion of the these molecules from the square (σ-aromatic) structure to the rectangular (π-antiaromatic) architecture. For the dianion $Al_4Li_4^{2-}$, the σ-equalization prevails over the π-distortion in $Al_4Li_4$ and for the dication $Al_4Li_4^{2+}$, π-equalization is the driving force for the square symmetric structure.